\documentclass[12 pt,a4paper]{article}
\usepackage[dvips]{graphicx}
\usepackage{graphicx}
\usepackage{amsmath,amssymb}
\usepackage{epsfig}
\usepackage{float}
\usepackage{fancybox}
\usepackage{pst-all}
\usepackage{natbib}

\newcommand{\be}{\begin{equation}}
\newcommand{\bfig}{\begin{figure}}
\newcommand{\ee}{\end{equation}}
\newcommand{\efig}{\end{figure}}
\newcommand{\bi}{\begin{itemize}}
\newcommand{\ei}{\end{itemize}}
\newcommand{\bear}{\begin{eqnarray}}
\newcommand{\eear}{\end{eqnarray}}
\newcommand{\ba}{\begin{array}}

\newcommand{\ea}{\end{array}}

\title{ Trading strategies in the Italian interbank market }
\author{
Giulia Iori\footnote{Corresponding author}  \\
{\small Department of Economics, City University}\\
{\small Northampton Square, EC1V 0HB London, U.K.} \\
{\small E-mail: g.iori@city.ac.uk} \\ 
Roberto Ren\`o\\
{\small Dipartimento di Economia Politica, Universit\'a di Siena}\\
{\small Piazza S. Francesco 7, 53100 Siena, Italy} \\
{\small E-mail: reno@unisi.it}\\
 Giulia De Masi  \\
{\small Physics Department, University of L'Aquila, Italy}\\
{\small Via Vetoio, 67010 Coppito, L'Aquila, Italy} \\
{\small E-mail: giulia.demasi@aquila.infn.it}\\
Guido Caldarelli\\
{\small INFM-CNR SMC Centre and Department of Physics, 
University of "La Sapienza"}\\
{\small  P.le A. Moro 2, 00185 Roma, Italy} \\
{\small E-mail: Guido.Caldarelli@roma1.infn.it}
}
 \date{\today}

\begin{document}
\maketitle
\begin{abstract}
Using a data set which includes all transactions among banks in the
Italian money market, we study their trading strategies and the
dependence among them. We use the Fourier method to compute the
variance-covariance matrix of trading strategies. Our results indicate
that well defined patterns arise. Two main communities
of banks, which can be coarsely identified as small and large banks, emerge.
\end{abstract}

\newpage
\section{Introduction}

Credit institutions in the Euro area are required to hold minimum
reserve balances within National Central Banks.
Reserves provide a buffer against unexpected liquidity shocks,
 mitigating the related fluctuations of market interest rates.
These reserves are remunerated at the main refinancing rate. In the
period under investigation, they had to be  
fulfilled only on average over a one-month maintenance period
 that runs from the 24th of a month to the 23rd of the following month, the end of maintenance day (henceforth EOM).
Banks can exchange reserves  on the interbank market with the
objective to minimize the reserve implicit costs. In Italy, exchanges
are regulated in the e-MID market.  

The objective of this paper is to analyze correlations in the  liquidity 
management strategies among banks in Italy, using an unique data set 
of  transactions with overnight maturity. The information includes
transaction prices, volumes and the encoded identity of quoting and
ordering banks. Thus we are able to disentangle the trading strategy
of each bank. 
There are indications that
not all credit institutions actively manage their minimum reserves.
Some institutions, typically smaller, tend to keep their
reserve account at the required level constantly
through the maintenance period. 

We adopt recently developed statistical techniques to reliably
measure correlations  
between  trading strategies, 
as proxied by the cumulative trading volume. The time series
are highly asynchronous, but the adopted methodology, namely the
Fourier method, is suitable to deal with this situation. We estimate
the variance-covariance matrix, and we analyze it using two
techniques: standard principal component analysis and network
analysis, with the latter providing information on the
presence of communities.  

We show that the spectrum of the variance-covariance matrix displays
only few eigenvectors which are not in agreement with the random
matrix prediction. In particular we find that the largest eigenvalue,
which reflects the total aggregation level of the strategies,
decreases as we approach the EOM date. The network analysis reveals
the existence of two main communities.

\section{Estimation of variance-covariance matrix}

To analyze the presence of   common factors in trading
strategies among different banks we estimate the
variance-covariance matrix of signed trading volumes. 

The Italian electronic broker market MID (Market for Interbank
Deposits)  covers  the whole existing domestic overnight deposit
market in Italy. 
This market is unique in the Euro area in being a screen based  fully
electronic 
 interbank market. Outside Italy interbank trades are largely
 bilateral or undertaken via voice brokers. 
Both Italian banks and foreign banks can exchange funds in this
market. 
The participating banks 
were 215 in 1999, 196 in 2000, 183 in 2001 and 177 in 2002.
Banks names 
are visible next to their quotes to facilitate credit line checking. 
Quotes can be submitted and a transaction is finalized if  the
ordering bank accepts a listed bid/offer. Each quote  is identified as
an offer or a bid.   
 An offer indicates that the transaction has been concluded at the
 selling price of the quoting bank  while a bid indicates that  a
 transaction has been concluded at the buying price of the quoting
 bank. 
  
Our data set consists of all the overnight transactions  concluded on
the e-MIB from January 1999 to December 2002 for a total of 586,007
transactions. For each contract we have information about the date and
time of the trade, the quantity, the interest rate and the encoded
name of the quoting and ordering bank.   

Our sample consists of $N_b=265$ banks, and our data span four trading
years.  In the analysis, our main results refer to a sub-sample of 85 banks who trade at least $900$ days (from the first
transaction to the last),  and with a number of transactions larger than $1000$. For comparison, in some cases, we also present the results of the analysis  for the all sample (in this last case the results are statistically less accurate).

Trading is highly asynchronous. While some banks trade as frequently as
every few minutes others can be inactive for several days.

\begin{figure}
\begin{center}
   \includegraphics [scale=0.3]{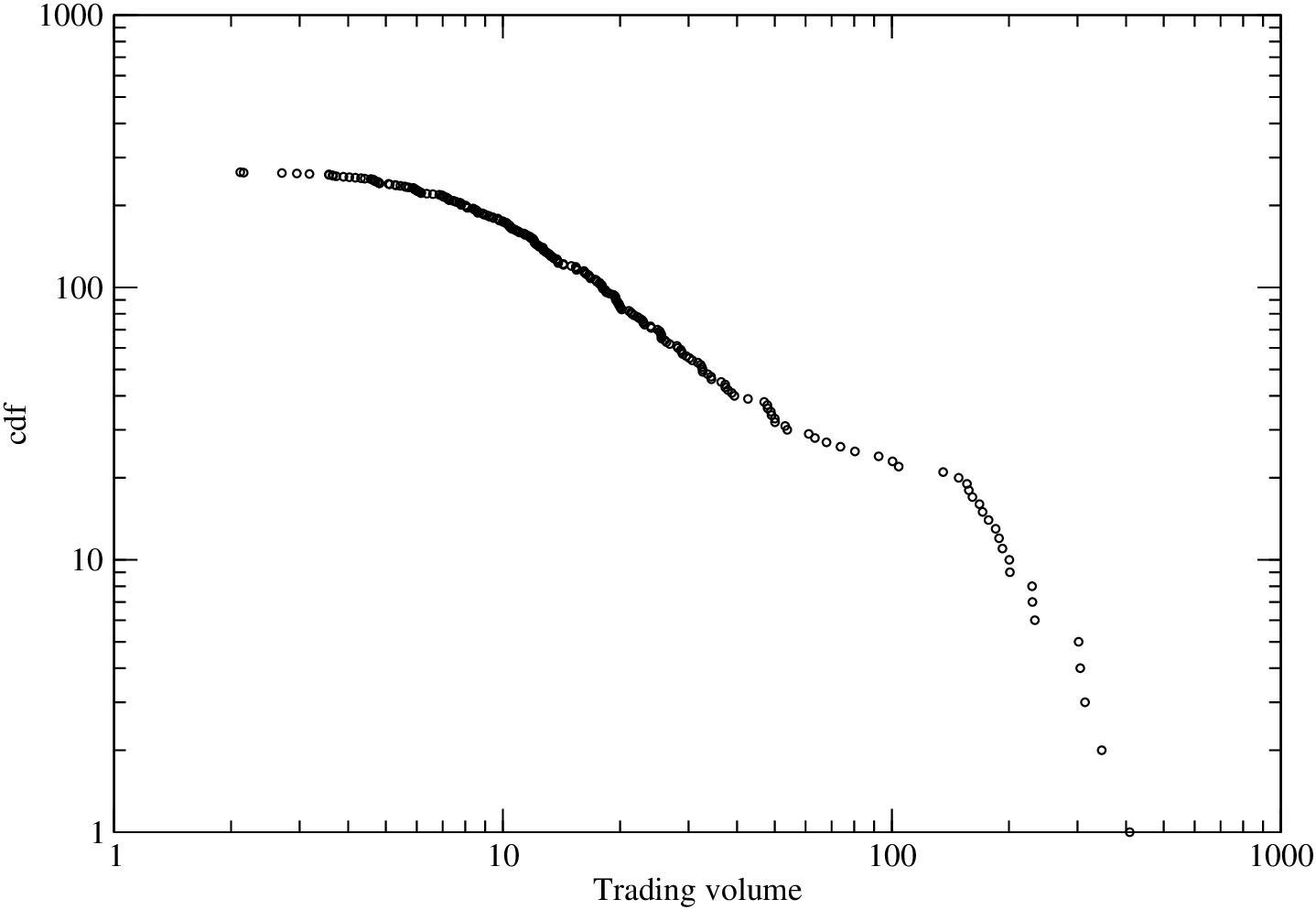}\hskip .5cm
   \includegraphics [scale=0.3]{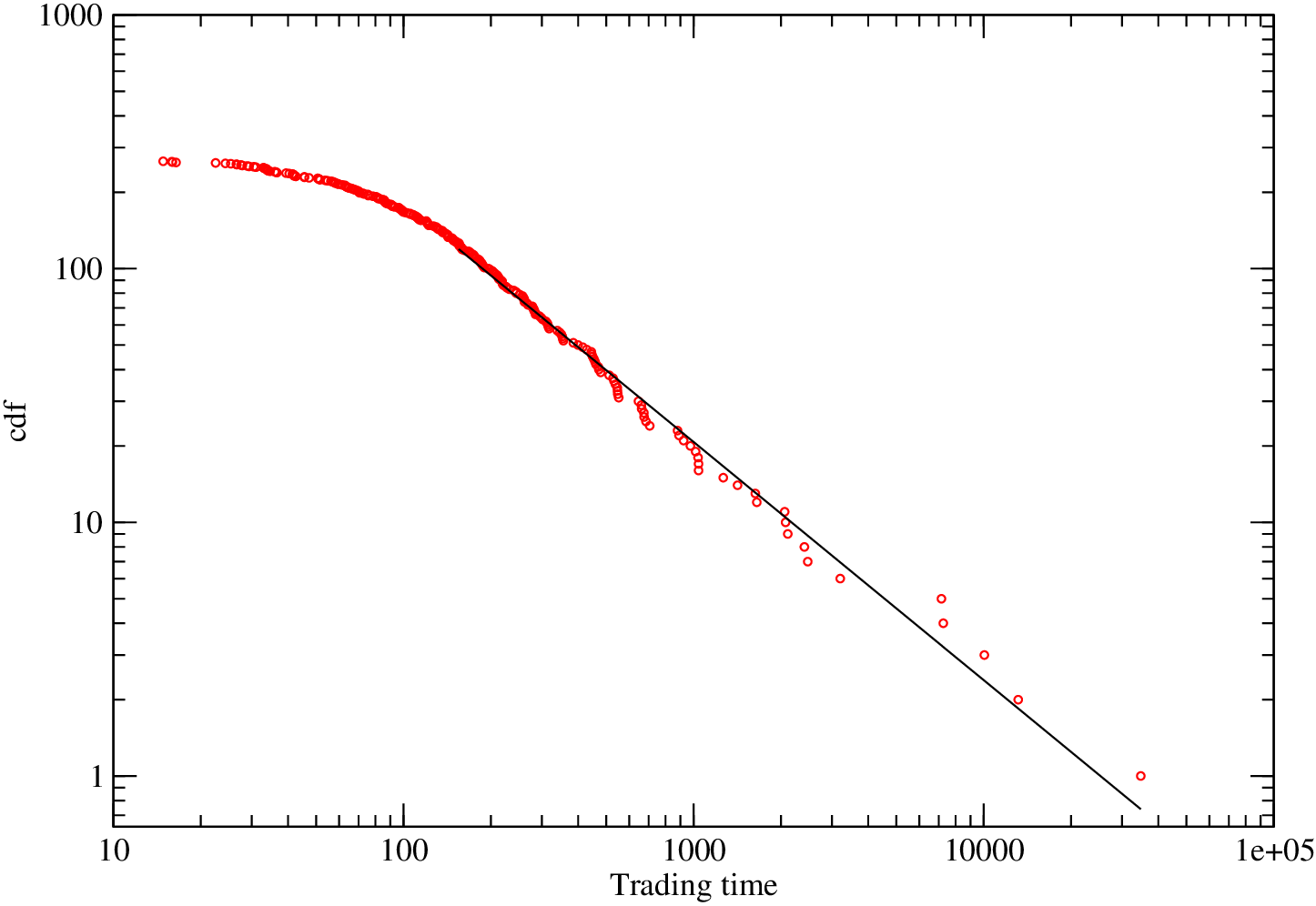}\vskip .2cm
    \caption{Cumulative distribution of average trading size  (left) and   waiting time  in minutes (right) across banks.  }
     \label{figure1}
    \end{center}
   \end{figure}

In figure \ref{figure1}
we plot the cumulative distribution of average volumes  (left) and
average waiting times (right)  across banks.  The distribution of
average waiting 
times is power law, revealing that there is not a typical scale for the
trading frequency in the system. 

\begin{figure}
\begin{center}
   \includegraphics [scale=0.4]{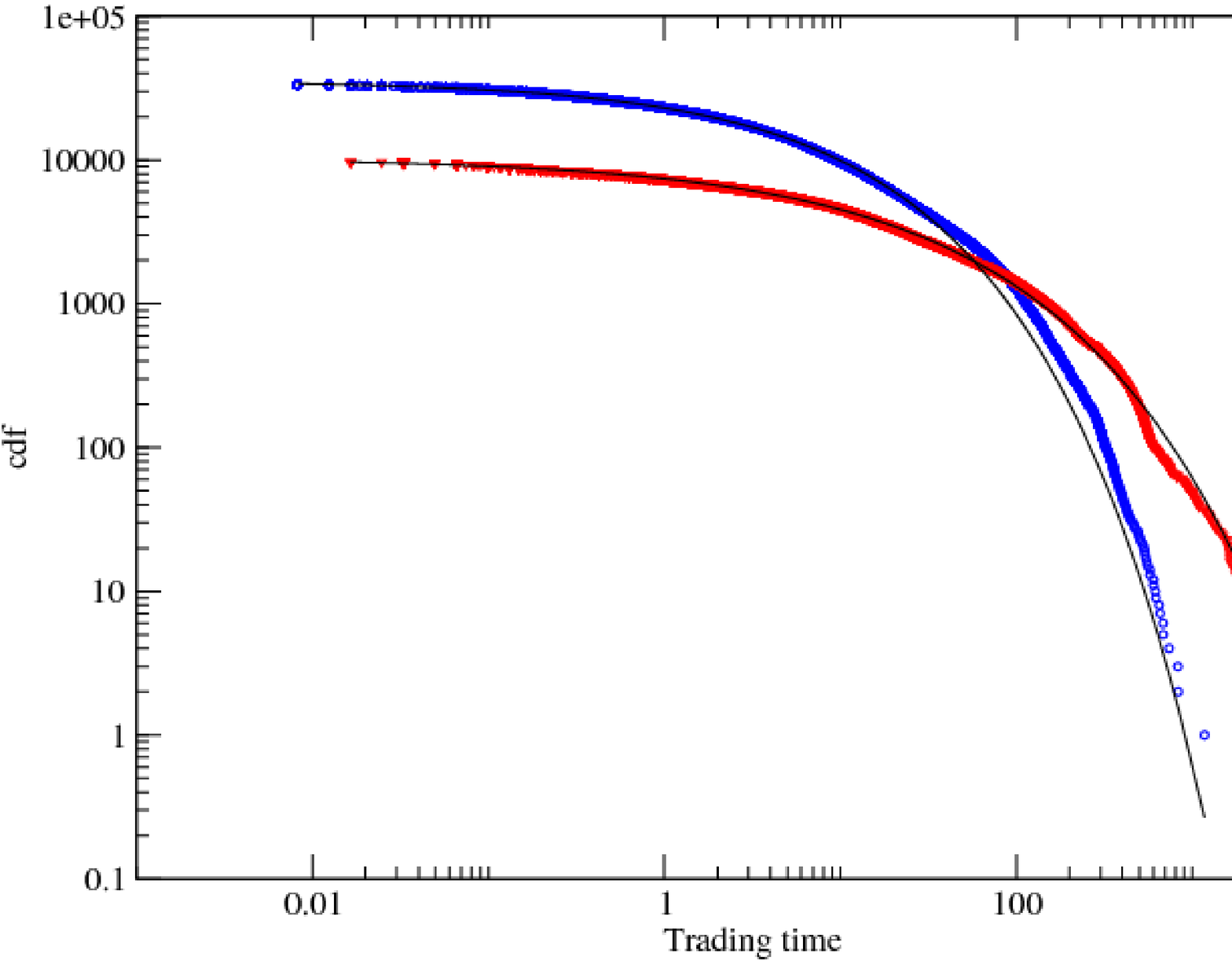}\hskip .5cm
    \caption{Distribution of  waiting time  for two banks and their
corresponding stretched exponential fit. The parameters of the fit are
reported in table \ref{table}.}
     \label{figure2}
    \end{center}
   \end{figure}

 In figure \ref{figure2}  we plot the distribution of waiting  times,
over the four year periods,  for two highly active banks.   
We find that the distribution follows a stretched exponential of the
form $a \exp(-t/\tau)^\beta$ [1].  
The parameters of the fit for the two banks 
are reported in table \ref{table}.

We denote the signed trading volume as  the cumulative volume of transactions
for a single bank, where every transaction is added with a plus sign
if the transaction is a sell, and with a minus sign if the transaction
is a buy. As an example, if a bank starts with a given liquidity,
lends some money and then borrows it to restore the initial liquidity,
the total signed trading volume will be positive and increasing in the
beginning, then decreasing to zero thereafter.

\begin{table}[b!]
\begin{center}
\begin{tabular}{|c|ccc|c|}
\hline 

Bank code & $a$ & $\tau$ & $\beta$ & $<dT>$ (minutes) \\
\hline
1 & 35589 & 5.90 & 0.466 & 16 \\
70 & 10315 & 16.14 & 0.396 & 27 \\
\hline
\end{tabular}
\caption{Parameter estimation for the cumulative distribution function
of waiting times for two highly active banks. The distribution is
fitted to a stretched exponential of the form
$a\exp(-t/\tau)^\beta$. $<dT>$ is the average waiting time between
transactions for the two banks.}
\label{table}
\end{center}
\end{table} 

Hence we compute the  signed trading volume $\bar{V}_t^i$, where the 
superscript $i$ denotes the bank, and we analyze their correlation matrix.
Figure 3 shows the signed trading volume time series for a
number of banks. It is apparent that there are some banks that follow
correlated strategies and others that follow anticorrelated
strategies. A structural break clearly appears in July 2001, when some
banks inverted their trading behavior. This change of behavior can
be associated with two events, at least. 
First, the official and market interest
rates of the Euro area changed their trend from positive to negative
at the beginning of 2001. Money market rates started increasingly to
price in a reduction in 
the ECB interest rates of 25 basis points that was eventually
decided on 30 August 2001. 
Furthermore the amount of liquidity provided by the 
European Central Bank increased in the summer 2001 to support economic growth.

\begin{figure}
\begin{center}
                  \includegraphics [scale=.7]{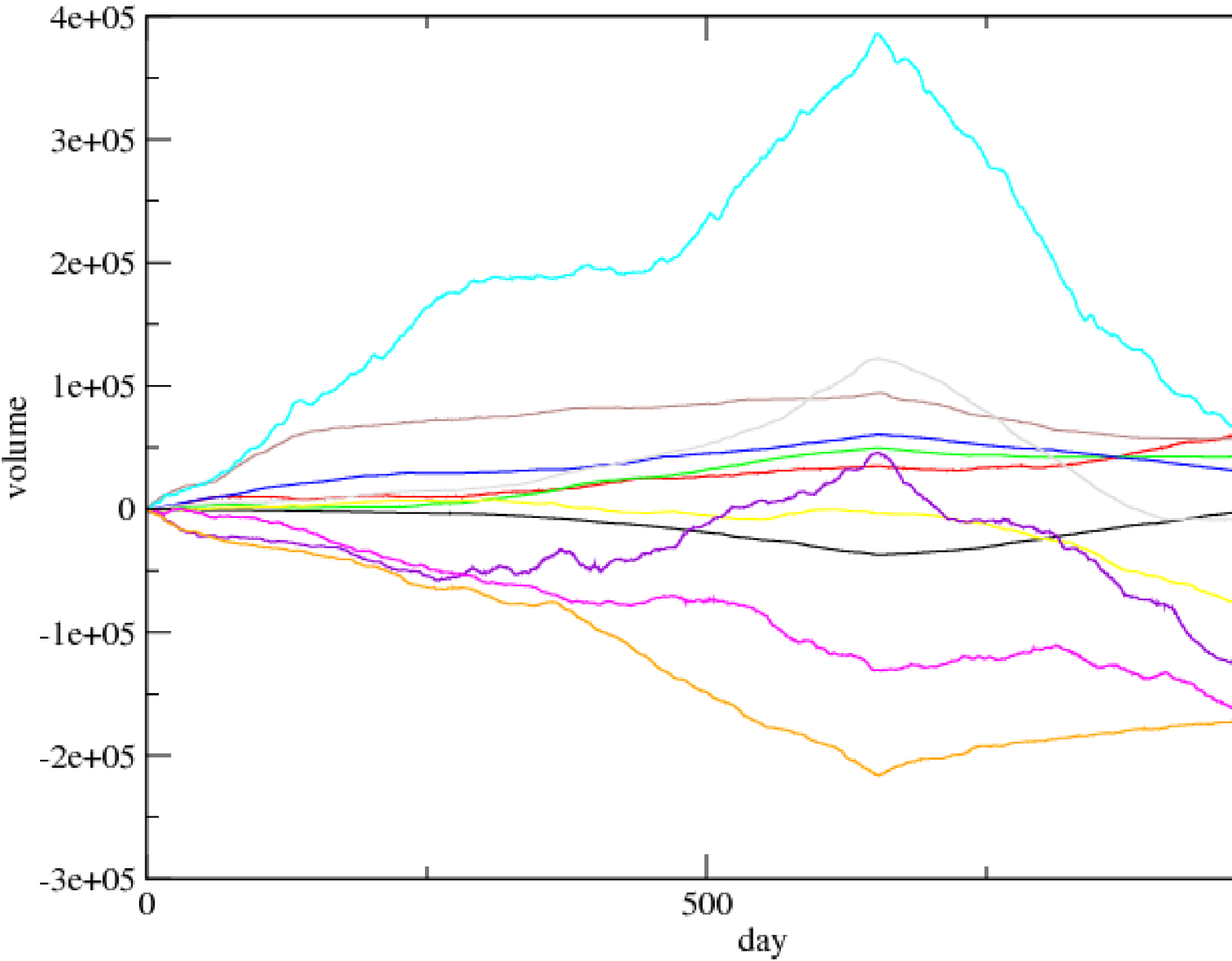}\vskip .2cm
    \caption{Sample of banks' strategies.}
    \end{center}
 \label{volume}
   \end{figure}
 
There are many issues in computing the variance-covariance matrix. 
First, we do not observe changes of signed volume in
continuous time, but only in form of discrete changes which occur when
a transaction between two banks takes place. Second, the transactions
do not occur  at the same time. Thus, given two banks, the changes in signed
volume are not synchronous. Furthermore some banks can trade more often
than others and we observe a wide spectrum of trading frequencies.
These difficulties make the implementation of a classical
Pearson-like variance-covariance estimator problematic.

To circumvent these difficulties, we adopt the Fourier estimator of [2]
 instead. This is based on the following idea. Take
a number of trajectories $X_i$ observed in $[0,2\pi]$, and compute the
Fourier coefficients $a_k(dX_i),b_k(dX_i)$ of $dX_i$. Then, the
covariance between $X_i$ and $X_j$ is estimated via:
\begin{equation}\label{covariance}
Cov(X_i,X_j) \simeq \frac{1}{M}\sum_{k=1}^M \left[a_k(dX_i)a_k(dX_j)+b_k(dX_i)b_k(dX_j)\right]
\end{equation} 
where $M$ is the largest frequency employed in computation which has
to be suitably selected. The actual covariance is estimated in the
limit $M\rightarrow\infty$.
The methodology is model-free, and it 
produces very accurate, smooth estimates. Most importantly, 
it handles the time series in their original form without any need of 
imputation or data discarding.  The estimator is based on integration
of the time series to compute the Fourier coefficients, thus it is
well suited to the uneven structure of the data. Moreover, it has a
natural interpretation in the frequency domain, which is exactly what
we aim to take care of, given banks'  different trading
frequencies. This estimator has been shown to perform much better 
than the Pearson estimator in this kind of situations
[3],[4]. It also performs well in  estimating
univariate volatility 
in the presence of microstructure noise[5]. 
 Typically, it has been used for
financial markets asset prices, e.g. on foreign exchange rates [6][7], 
stock prices [8] and stock
index futures prices [9,10,11].
In this paper, we use it to
analyze the cross-correlation among cumulative volumes.
We refer to the quoted papers for the
description of the implementation.

When computing covariances, we have to select carefully the maximum
frequency employed in the computation. When the frequency
increases we observe the Epps effect [12], that is the
absolute value of 
the correlation is biased toward zero, see [3]. This is
evident in Figure 4, where we show (left) the correlation as a function of
the frequency for two given bank trading strategies, and (right) the
average  positive and negative correlation among all banks. It is important to
remark that, given the definition of the trading strategy $\bar{V}_i$,
whenever a bank increases its cumulative volume, there is a bank which
decreases its cumulative volume, namely the bank who traded with it. 
Thus, negative correlations among trading strategies arise naturally.
Since the less active bank in our sample makes about 1,000 transactions, we
can choose a maximum cut-off frequency $M=500$, where $M$ is the
largest Fourier harmonic used in (\ref{covariance}). 
We compare two different frequencies, $M=100$, corresponding to
a time scale of nearly $10$ days, and $M=500$, corresponding to a time
scale of a couple of days. The largest the frequency the smallest the  error and the largest the bias toward zero. 
Figure 5, shows the 
distribution of the correlations among trading strategies for $M=100$
and $M=500$.  Correlations with $M=500$ are more centered around zero,
because  of the Epps effect, nonetheless  there is not a great
difference between the two, indicating that  the bias is not so
relevant.
Thus the  following results are obtained using $M=500$.

\section{Principal Component Analysis}

We analyze the correlation matrix of the trading
strategies with the technique of random matrix theory (RMT), 
in line with the work of [13,14] on
stock prices. Random matrix properties are derived in [15].
We find that at least two eigenvalues do not fit the predictions of
RMT, see the inset in Figure 6.
The economic interpretation of this fact is quite straightforward.
 If a bank lends and borrows
over time in a non strategic way, the correlation among trading
strategies should conform to the predictions of RMT. The fact that
this is not the case means that banks do not behave randomly but a certain level of coordination can be observed. 
[16]  for example have shown   that  over 2002 small
banks have overall been acting as lenders,  while larger banks have
overall been acting as borrowers. 

We finally check whether there is some deterministic pattern in the
evolution of  
trading strategies over the maintenance period. To this purpose we
compute the daily 
correlation matrices, and average those which are at the same distance
from the EOM. 
It is well known that the behavior of banks is different
near the EOM  [16,17,18]

We find that the first eigenvalue has a decreasing explanatory power over
the maintenance period, see Figure \ref{eigen12}. 
On the contrary, the impact of the second largest eigenvalue, shown in
the inset of figure \ref{eigen12}, is constant over the maintenance period.

In the very first day of the maintenance period there is no distinct
coordination among the strategies. Coordination increases and  then
it declines gradually in the last few days of the maintenance period. The
interpretation of this 
empirical result is quite straightforward. When the EOM
day is approaching, banks take more care in fulfilling their reserve
obligations than in pursuing their preferred strategy (being a lender
or a borrower). Thus, they transact more for pure liquidity reasons,
and the correlation among trading strategies is less strong.

%
%
\begin{figure}
\begin{center}
   \includegraphics [scale=0.25]{epps_bank_1_47.eps}\hskip .5cm
   \includegraphics [scale=0.25]{epps_average.eps}\vskip .2cm
    \caption{Epps effect in the correlations of two individual banks, bank 1 and bank47 (left) and  average Epps effect  (right). }
    \end{center}
 \label{figure3}
   \end{figure}
   
\begin{figure}
\begin{center}
   \includegraphics [scale=0.5]{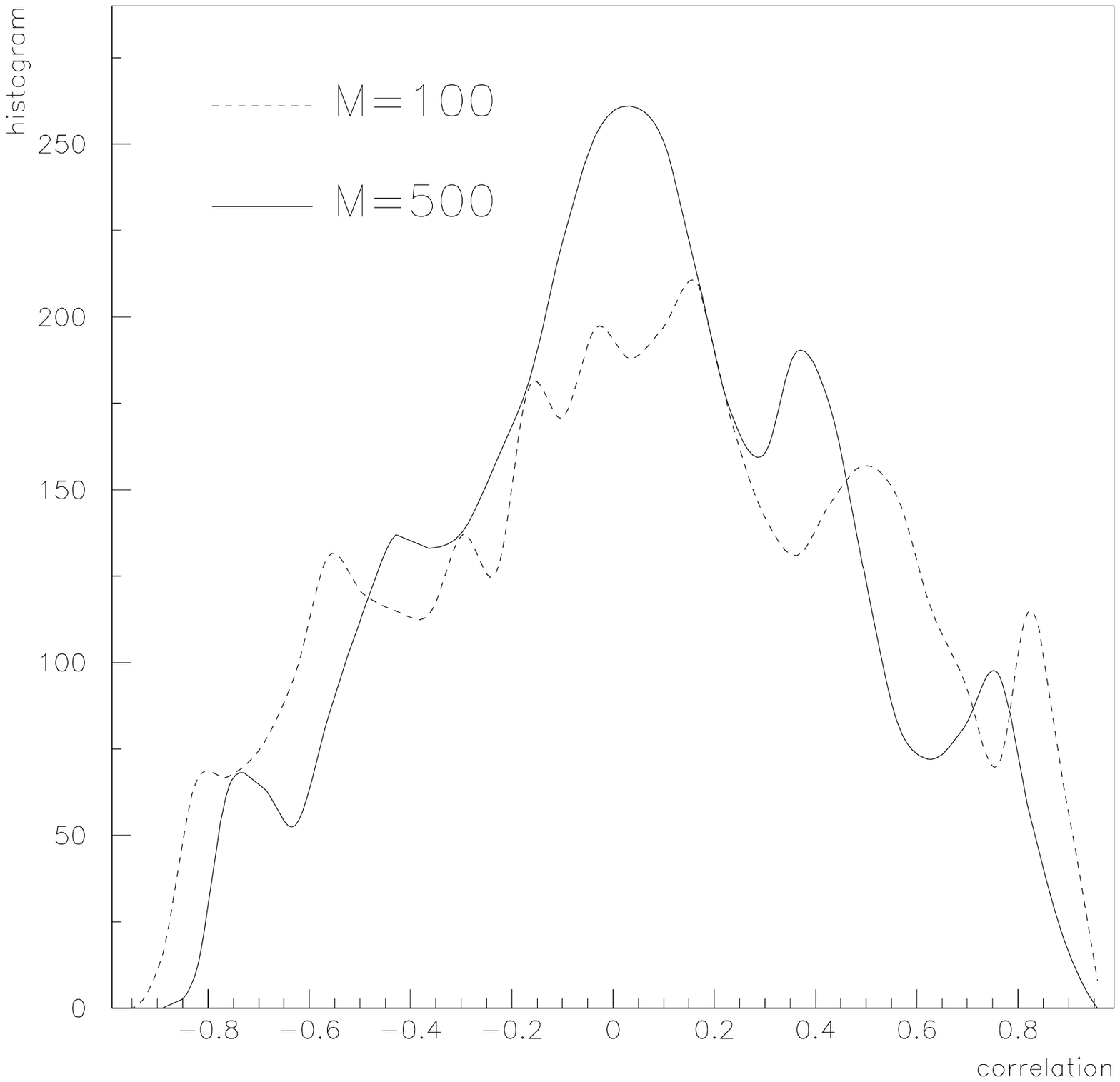}
    \caption{Correlation distribution at two different cut-off frequencies. }
    \end{center}
 \label{figure4}
   \end{figure}
   
   \begin{figure}
\begin{center}
     \includegraphics [scale=0.5]{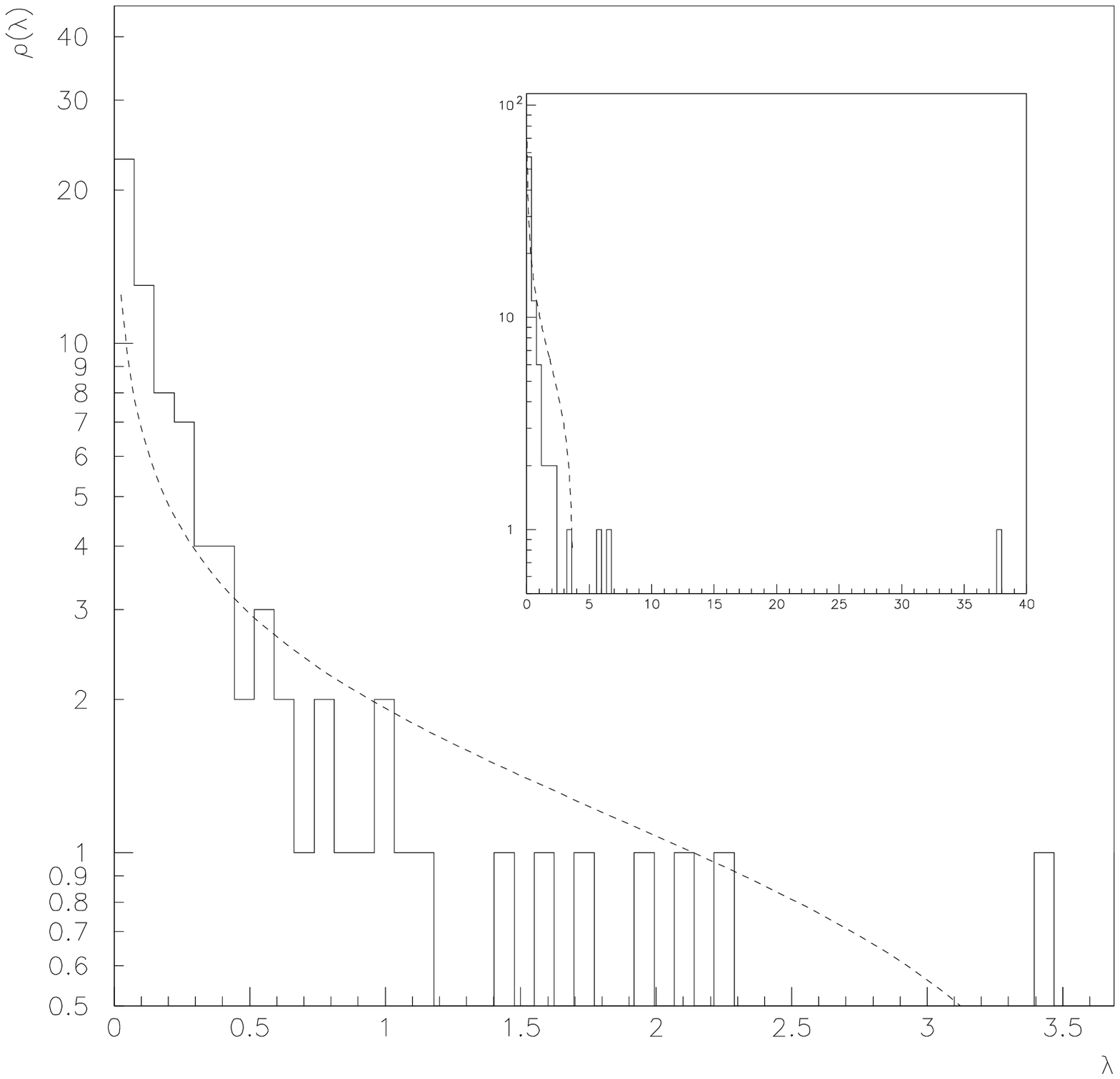}\vskip .2cm
     \caption{Eigenvalues distribution at $M=500$ for the smallest
eigenvalues (histogram) together with the prediction of RMT (dashed
line). 
In the
inset,  the whole distribution is shown. }
    \end{center}
 \label{figure5}
   \end{figure}

%
   
\begin{figure}
\begin{center}
   \includegraphics [scale=0.55]{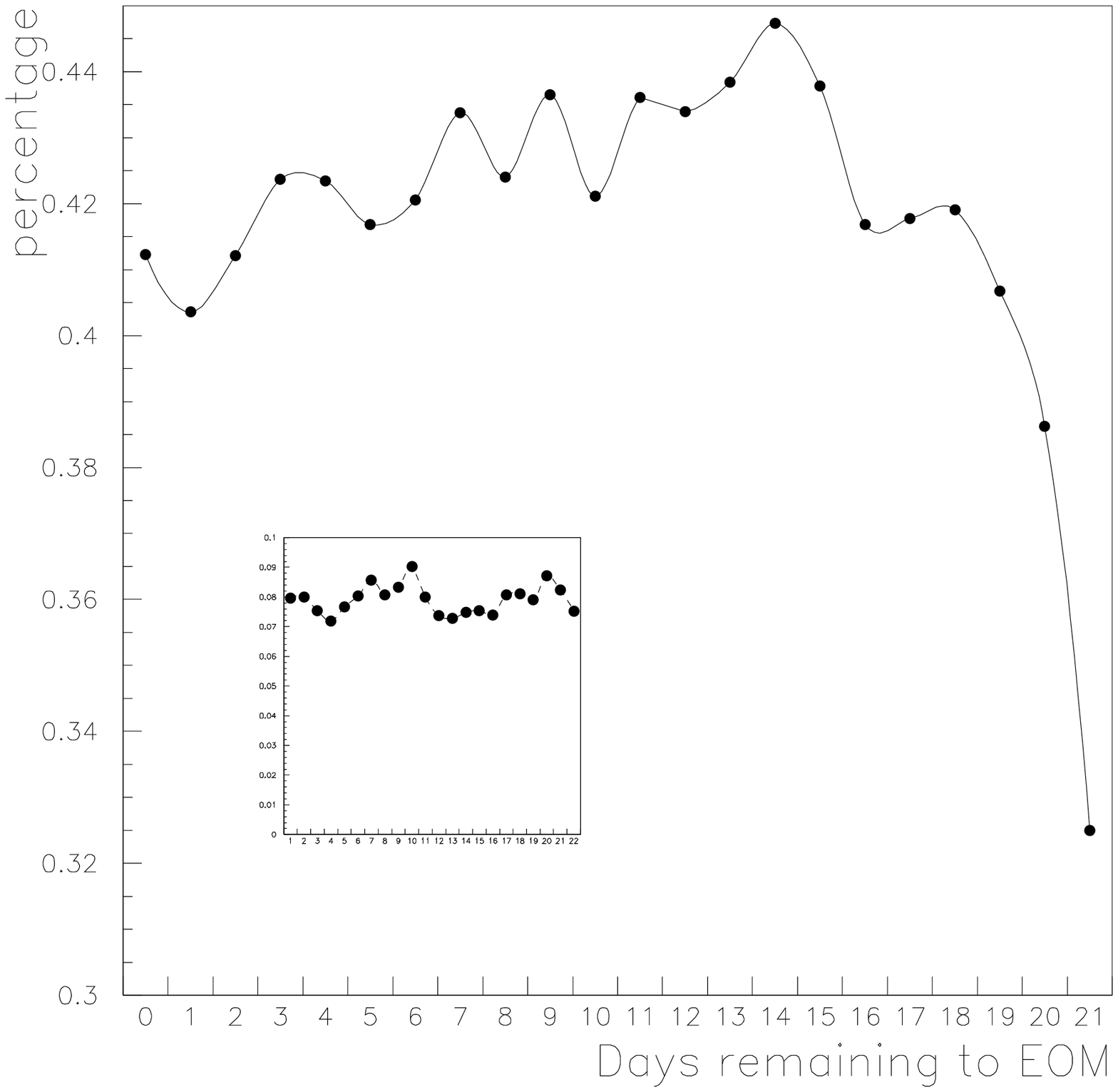}\vskip .2cm
    \caption{Explanatory power of largest eigenvalue as a function of the distance from EOM day. In the inset  the same plot for the  second largest  eigenvalue }
    \label{eigen12}
    \end{center}
   \end{figure}
   
%

\section{Communities structure}

The aim of this section is to identify groups, or communities,  of
banks with similar trading strategies. To help visualizing the results
we use the following taxonomy. We represent large Italian banks with
blue circle,  foreign banks with red circles, medium Italian banks
with green and small banks with yellow. The two banks labeled in pink
are two central institute of categories. 
 
\begin{figure}
\begin{center}
          \includegraphics [scale=1]{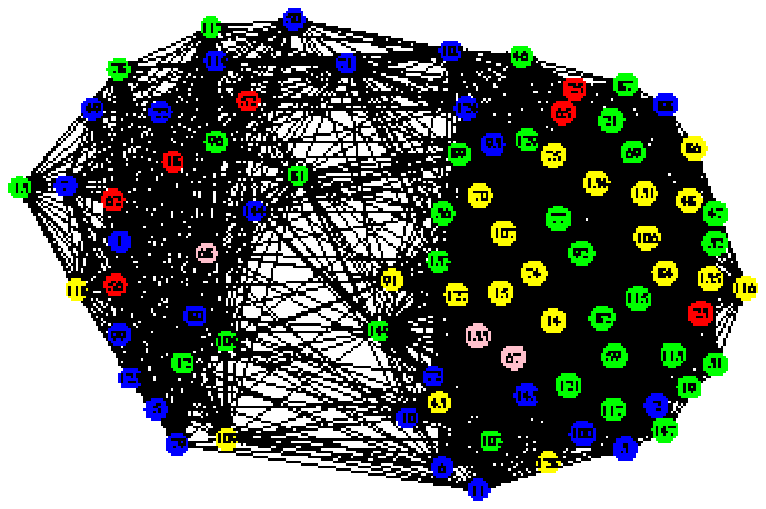}  \vskip 2cm
    \caption{Positive correlation graph.}
 \label{figure7}
    \end{center}
   \end{figure}

\begin{figure}
\begin{center}
 \includegraphics [scale=1.6, angle=-30]{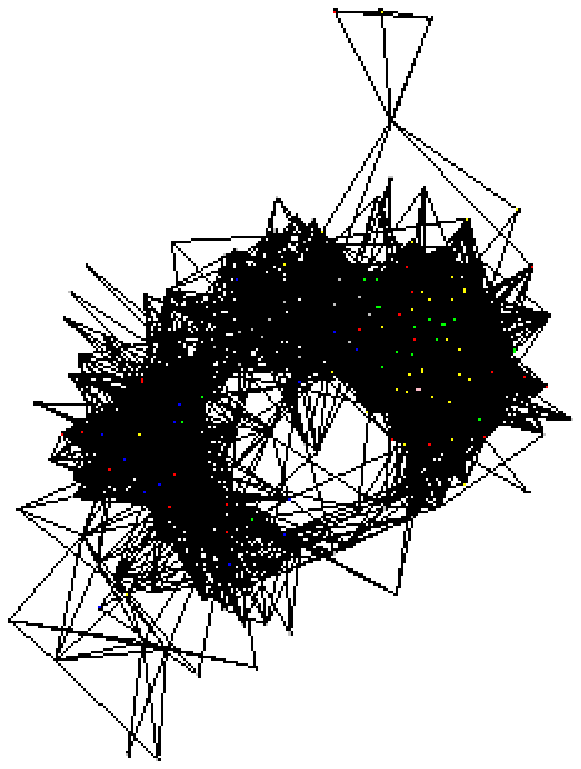}  
    \vskip -1cm     \includegraphics [scale=.7]{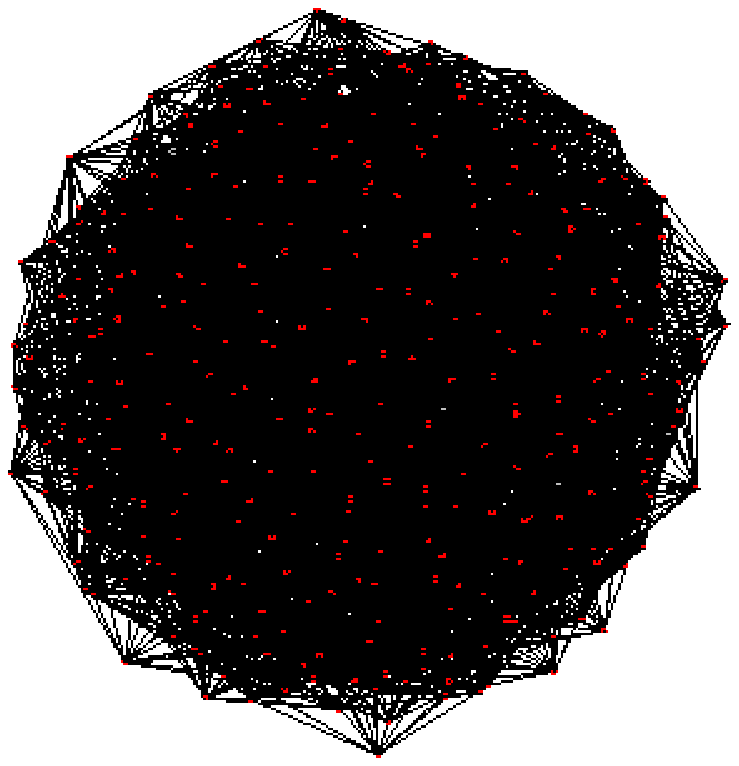}     \vskip 2cm        
    \caption{Positive  correlations graph for all banks and M=25, for original transactions (top) and reshuffled transactions (bottom).}
 \label{figure7bis}
    \end{center}
   \end{figure}

A simple way to identify communities in the system is to plot the
correlation matrix as a graph\footnote{For visualizations we used
  Graphviz Version 1.13 with  the Energy Minimized layout
  [19]}, where banks are the vertices and links among
them exist their strategies are positively correlated. Figure
\ref{figure7}  (left) clearly evidences  two separate communities of banks.
To test that the two communities are not present only when selecting the frequency $M=500$ we have repeated the calculations by changing $M$. The results are consistent at all frequencies. For example   in figure \ref{figure7bis} (top) we plot the results obtained by  selecting a lower frequency $M=25$, which correspond to a time scale of about one month and including all banks that  trade at least once a month. 
Trading of bank reserves at different frequencies is driven in principle  by
different consideration. On an intra-day level, the main
   determinant is to target overnight balance without
   exceeding exposure limits,  while on a monthly level, the aim is to meet
   reserve requirements. Movements in reserves at a lower frequency are
   mainly determined by the developments in banks' other balance sheet
   positions over which banks have little control so  that banks may not be able to play strategically at low frequencies but they may be acting strategically over longer time scales.  Nonetheless we do not observe a clear difference in the correlation matrices over the two time scales and we still observe the  same two communities in the correlation network in figure \ref{figure7bis} (top).
 
To check that the method does not introduce spurious correlations and that communities do not emerge purely from the fact that the in the market there are simultaneously buyers and selles we have repeated the calculations of correlations by reshuffling the transactions (i.e. assigning each transaction to a random buyer and a random seller). The resulting graph is plotted in Figure \ref{figure7bis} (bottom). In this case clearly no comunities emerge.

\begin{figure}
\begin{center}
                  \includegraphics [scale=2.5]{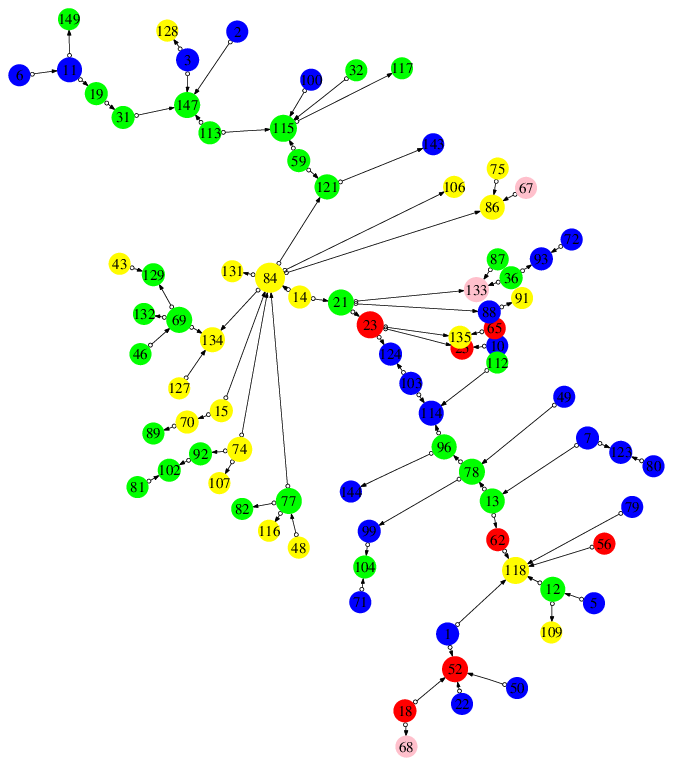}\vskip .2cm
    \caption{MST of correlations.}
 \label{figure8}
    \end{center}
   \end{figure}

\begin{figure}
\begin{center}
   \includegraphics [scale=1.0]{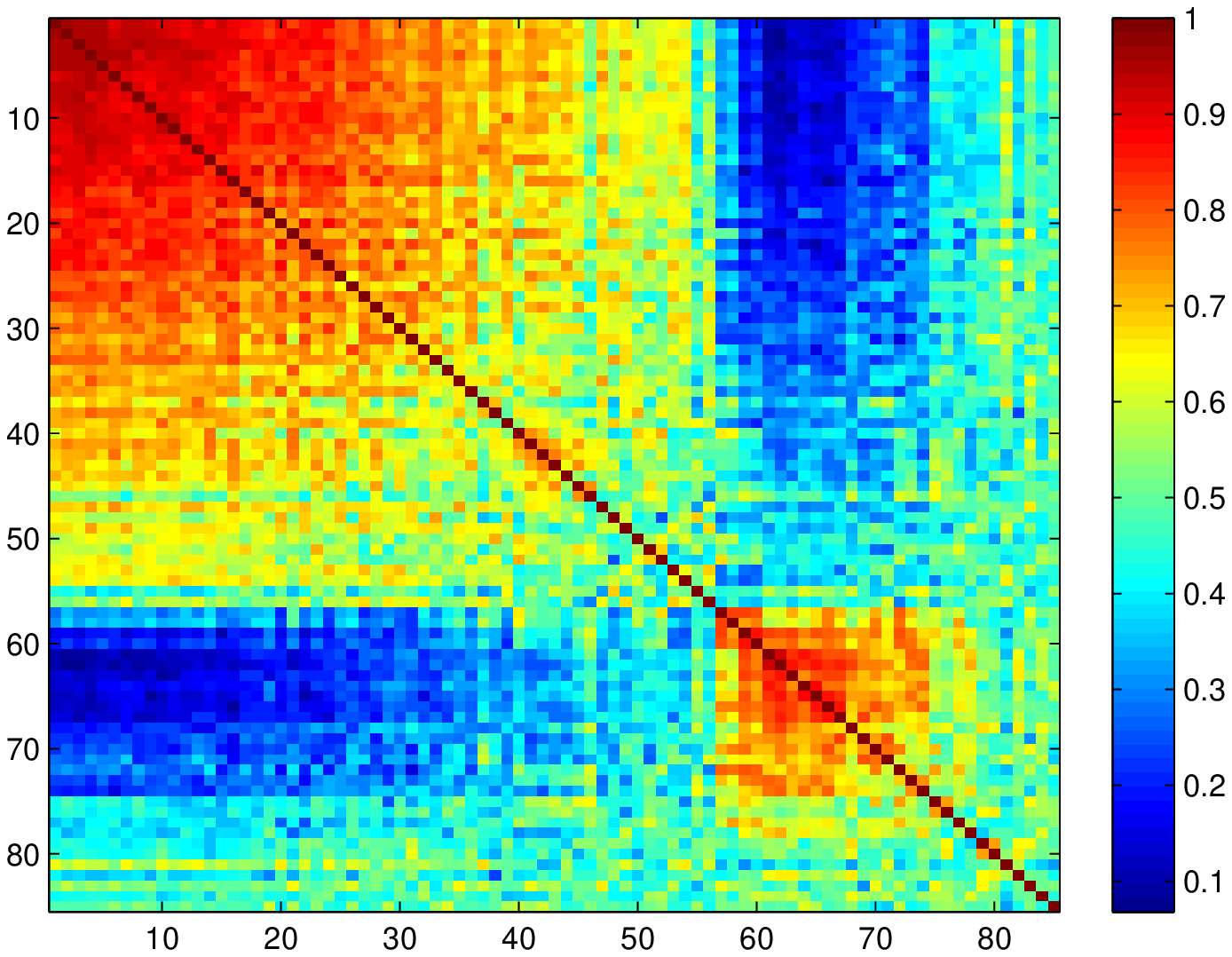}\vskip .2cm
    \caption{Correlation matrix. Banks  are ordered according to  the hierarchical tree of figure \ref{figure8}.}
 \label{figure9}
    \end{center}
   \end{figure}

In figure \ref{figure8} we plot the minimum spanning tree generated by
the correlation matrix using the definition of distance given in
[20,21]. 
On this tree, we can identify the same  two communities identified
before. The branches departing  from bank number 103 are exactly the
two groups on the left and the right of figure \ref{figure7}.   
In figure \ref{figure9} we plot the overall correlation matrix where
banks are ordered
according to  the hierarchical tree of figure
\ref{figure8}, as in [22]. Again two distinct groups
appear. 

More sophisticated  techniques for detecting community structure have been
proposed in the last few years
[23,24,25,26,27,28].
Some of them  are based on the edge betweenness introduced by
[29], also known as the NG--algorithm. 
To detect communities with this method the algorithm
removes the edges with largest betweenness and in doing so it  
splits, step by step, the whole network into disconnected
components.  One problem with  this
method is that there is not an established 
criterion to stop the splitting process unless one knows, a priori, how
many communities there are.  
To overcome such problems, approaches based on the spectral
analysis have been recently adopted [30].  This approach
does not need any a priori knowledge about the number of
communities in the network (which is the most common case
in actual networks) but it is based on a constrained optimization
problem, in which we consider the nodes of the networks as objects
linked by an harmonic potential energy.

 \begin{figure}
\begin{center}
    \includegraphics  [scale=1.]{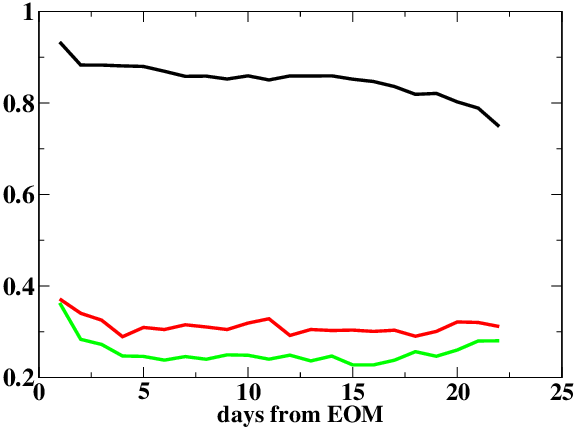}
  \caption{Distance from value 1 of the second, third and fourth eigenvalues for all days of the month: only the second eigenvalue is close to one indicating that only two communities are present. The  second eigenvalue is closer to one on the  EOM indicating that communities become  more pronounced.}
 \label{figure10}
    \end{center}
   \end{figure}

 \begin{figure}
\begin{center}
   \includegraphics [scale=.3, angle = 0]{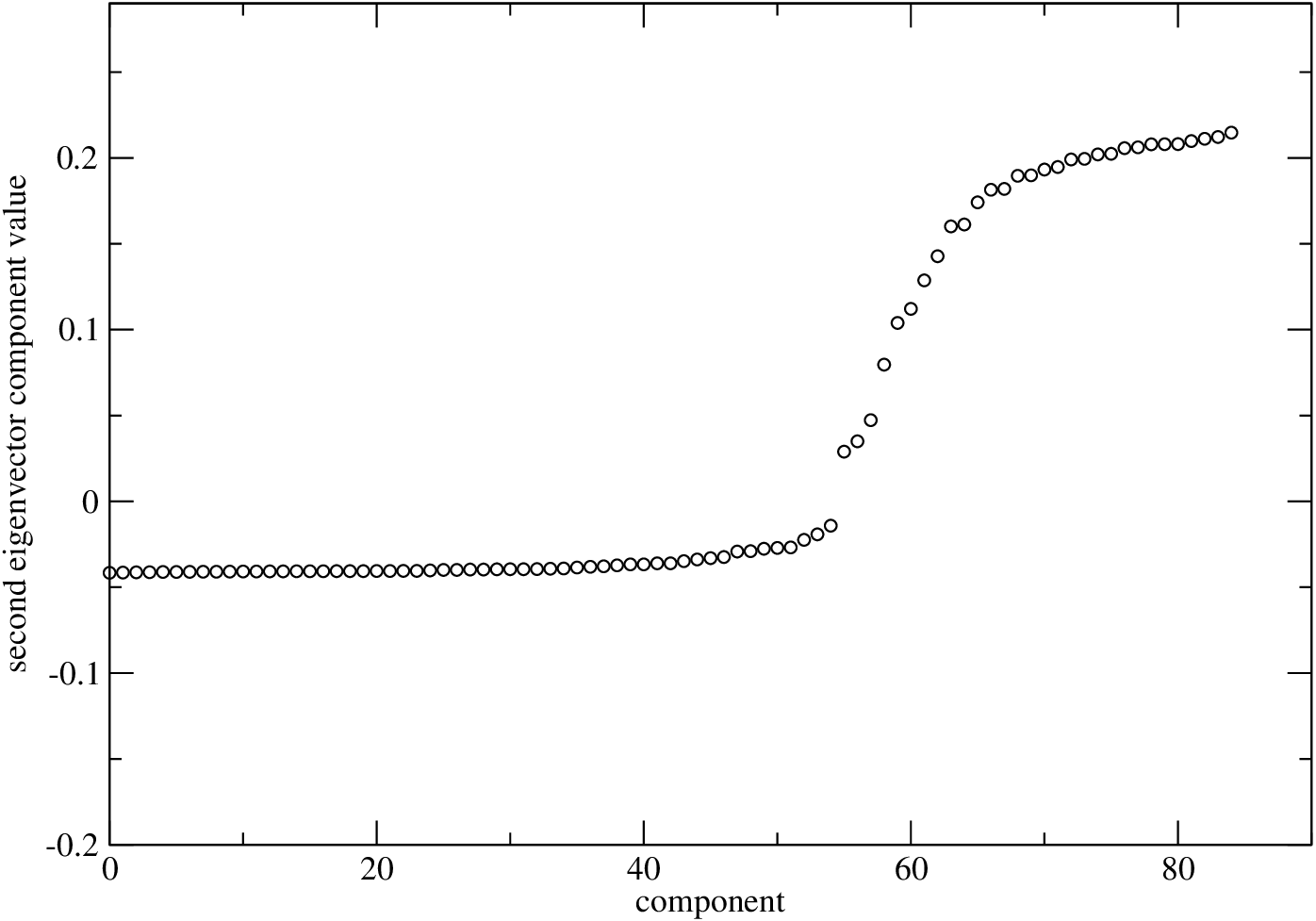}  \hskip .5cm 
     \includegraphics [scale=.3, angle = 0]{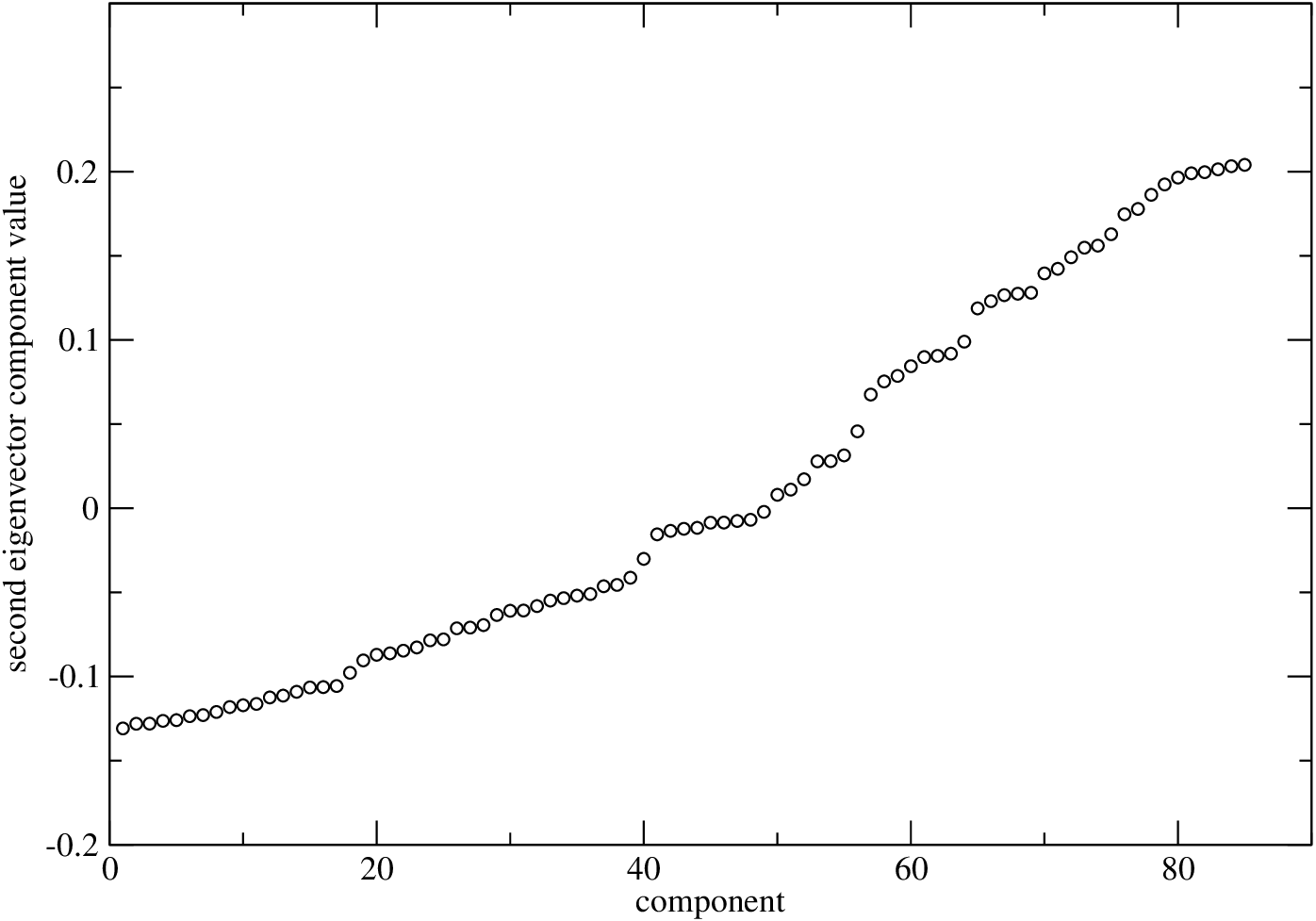}   
     \caption{(Left) Second eigenvector components: the presence of a
      step like profile reveals the presence of two communities in the
      network. (Right) Second eigenvector components for the reshuffled matrix.} 
 \label{figure11}
    \end{center}
   \end{figure}

\begin{table}
\begin{center}
\begin{tabular}{|c|c|c|} \hline
{\it Detected } & {\it First} & {\it Second} \\ 
{\it communties} & {\it Community} & {\it Community}  \\ 
  \hline
{} & {\it 113, 147, 115,} &{\it 114, 112, 81, 52,}\\
{} & {\it 23, 121, 21, 84,} & {\it 96, 91, 109, 80,}\\
{} & {\it  59, 82, 135, 2,} & {\it 22, 49, 78, 123,}\\
{} & {\it  86, 106, 47, 15,} & {\it 79, 56, 12, 7,}\\
{} & {\it  32, 134, 14, 107,} & {\it 13, 118, 62, 1}\\
{} & {\it  143, 131, 74, 67,} & {}\\
{} & {\it  100, 3, 116, 88,} & {}\\
{} & {\it  31, 87, 70, 92,} & {}\\
{} & {\it  77, 10, 133, 48,} & {}\\
{} & {\it  89, 71, 19, 36,} & {}\\
{} & {\it  128, 117, 50, 129,} & {}\\
{} & {\it  149, 69, 6, 127,} & {}\\
{} & {\it  124, 75, 102, 25,} &{}\\
{} & {\it  18, 46, 132, 103} & {}\\
 \hline 
\end{tabular}
\caption{Communities detected\label{communities tbl}}
\end{center}

\end{table}


Many spectral methods [31,32,33] 
are based on the analysis of  simple
functions of the (weighted) connectivity matrix $A$, 
identified here as the correlation matrix. In particular, the
functions of $A$ adopted are the Laplacian matrix $L=K-A$ and the
Normal matrix $N = K^{-1}A$, where $K$ is the diagonal matrix with
elements $k_{ii}=\sum_{j=1}^n a_{ij}$ and $n$ is the number of
nodes in the network. In most approaches, concerning undirected
networks, $A$ is assumed
 to be symmetric. 
The matrix $N$ has always the largest eigenvalue equal to one,
 due to row normalization.
 In a network with an apparent cluster structure, the matrix $N$
has also a certain number $m-1$ of eigenvalues close to one, where
$m$ is the number of well defined communities, the remaining
eigenvalues lying a gap away from one. The
eigenvectors associated to these first $m-1$ nontrivial
eigenvalues, also have a characteristic structure. Define $x_i$ as the
position if the $i$-th vertex after sorting vertices by the value
of its corresponding component in one eigenvector. The components
corresponding to nodes within the same cluster have very similar
values $x_i$, so that, as long as the partition is sufficiently
sharp, the profile of each eigenvector, sorted by components, is
step-like. The number of steps in the profile gives again the
number $m$ of communities.
Here we  find just one  eigenvalue close to one (see figure
\ref{figure10}), suggesting the existence of only two
communities. We analyze the component of the associated second
eigenvector in the left side of figure \ref{figure11}  which shows the
characteristic step like profile, identifying the  two communities. 
As shown in Table 2 the communities are the same as those in figure \ref{figure7}. The
group on the  top right in figure \ref{figure11} corresponds to the
group on the left in figure \ref{figure7}.
On the right side of figure \ref{figure11} we plot the second eigenvector components for the correlation matrix obtained by reshuffling the transactions. The step like shape is not visible in this case.

 As the second eigenvalue approaches  one,  the communities becomes better
defined. Figure \ref{figure10} shows  that near the EOM communities become more
pronounced. This is not in contrast with our previous finding that the
overall level of aggregation in the system decreases when
EOM approaches. In fact, the overall correlation may decrease while the
correlation of strategies  inside the same community can become
stronger.
The separation between the two communities is less pronounced far from
the EOM. 
We have also observed that very  few  banks change community during
the maintenance period.    
 
Finally we note that the first community  is composed predominantly of
large and foreign 
banks (red and blue circles if figure \ref{figure7}) while the second
community involves 
predominantly small banks (green and yellow circles  if figure
\ref{figure7}). Furthermore,  the
first community displays a pronounced geographical feature,  that is
all banks belonging to it are located in the northern part of Italy.   

\section{Conclusions}

We investigated the Italian segment of the European money market over
the period 1999-2002 using a unique data-set from which it is possible
to reconstruct the trading strategy of each participating bank. We
used the Fourier method to estimate the variance-covariance matrix
that we analyzed, since it is the most suitable to the unevenly
structure of the data.
We then analyzed the variance-covariance matrix using standard PCA and
tools borrowed from the analysis of complex networks.
We find that two main communities emerge, one mainly composed by large
and foreign banks, the other composed by small banks. Banks act
predominantly as borrowers or lenders respectively, with an inversion
of their behavior in July 2001. Moreover, 
the  analysis reveals that while overall trading
strategies becomes less correlated when the EOM approach,  the
communities become more pronounced on the EOM date. 

 \section*{Aknowledgements}
 We thankfully acknowledge financial support from the ESF--Cost Action P10 "Physics of Risk". 

\section*{References}

[1] E. Scalas, R.Gorenflo, H.Luckock, F.Mainardi, M.Mantelli, and M.Raberto, Quantiative Finance 4 (6) (2004) , 695-702.

[2] P. Malliavin and M. Mancino, Finance and Stochastics, 6 (1) (2002), 49-61.

[3] R. Reno', International Journal of Theoretical and Applied Finance,  6 (1) (2003), 87-102. 

[4] O. Precup, G. Iori, Physica A 344 (2004), 252--256.

[5] M.O. Nielsen and P.H. Frederiksen, Working paper, Cornell University. (2005).

[6] E. Barucci and R.Reno', Journal of International Financial Markets, Institutions and Money 12 (2002), 183-200

[7] E. Barucci and R.Reno', Economics Letters 74(2002), 371-378.

[8] M. Mancino and R. Reno', Applied Mathematical Finance 12 (2) (2005), 187-199.

[9] R. Reno' and R. Rizza, Physica A 322 (2003), 620-628.

[10] M. Pasquale and R. Reno', Physica A  346 (2005), 518-528.

[11] S. Bianco and R. Reno',  Journal of Futures Markets 26 (1) (2006), 61-84.

[12] T. Epps, Journal of the American Statistical Association 74 (1979),291-298.

[13] V. Plerou, P.Gopikrishnan, B.Rosenow, L.NunesAmaral, and E.Stanley, Physical Review Letters 83(7)(1999), 1471-1474.

[14] L. Laloux, P.Cizeau, J.-P. Bouchaud, and M.Potters , Physical Review Letters 83 (7) (1999), 1467-1470.

[15] Mehta, M., "Random matrices", Academic Press, New York (1995). 

[16] G. Iori, G.DeMasi, O.Precup, G.Gabbi, and G.Caldarelli, A network analysis of the Italian interbank money market, submitted paper (2005).

[17] P.Angelini,  Journal of Money, Credit and Banking 32, (2000) 54-73.

[18] E. Barucci, C.Impenna, and R.Reno', The Italian overnight market: microstructure effects, the martingale hyphotesis and the payment system, Temi di Discussione, Bank of Italy N. 475 (2003).

[19] T. Kamada and S.Kawai, An algorithm for drawing general inderected graphs, Information Processing Letters 31 (1)  (1989), 7-15.

[20] R.N. Mantegna, and H.E. Stanley, An introduction to econophysics, Cambridge University Press (2000)

[21] G. Bonanno, F.Lillo, and R.N. Mantegna, Quantitative Finance 1 (2001), 1-9.

[22] G. Bonanno, F.Lillo, and R.N. Mantegna, Physica A 299 (2001), 16-27.

[23] F. Radicchi, C.Castellano, F.Cecconi, V.Loreto, and D.Parisi, Proceedings of the National Academy of Science of USA, 101(9) (2004), 2658-2663.

[24] J. Reichardt and S.~Bornholdt Physical Review Letters 92 (21) (2004), 218701.

[25] A. Clauset, M.E.J. Newman, and C.Moore, Physical Review E 70 (2004), 066111.

[26] M. E.J. Newman, SIAM Review 45 (2) (2003), 167-256.

[27] J. Duch and A.Arenas, Physical Review E 72 (2005), 027104.

[28] L. Danon, A.Diaz-Guilera, J.Duch, and A.Arenas, Journal of Statistical Mechanics, P09008  (2005).

[29] M. Girvan, and M.E.J. Newman, Proceedings of the National Academy od Science of USA 99 (12) (2002), 7821.

[30] A. Capocci., V.D.P. Servedio, G.Caldarelli, and F.Colaiori, Physica A 352 (2-4) (2005), 669-676.

[31] K.M. Hall, Management Science 17(1970), 219.

[32] A.J. Seary and W.~D. Richard, Proceedings of the International Conference on Social Networks, 1: Methodology, Volume~47 (2005).

[33] J.M. Kleinberg, ACM Computing surveys 31(4es), 5 (1999).

\end{document}